\documentclass[onecolumn,showpacs,preprint,tightenlines,10pt]{revtex4}
\pdfoutput=1
\usepackage{graphicx,anysize}
\usepackage{bm,bbm,epstopdf}
\usepackage{multirow}
\usepackage{amsmath}
\usepackage{bbm,subfigure,ulem}
\usepackage{amssymb}
\usepackage{threeparttable}
\usepackage{float}
\usepackage{varioref,hyperref}
\hypersetup{colorlinks,citecolor=blue,linkcolor=blue,urlcolor=blue}
\usepackage{epic}
\usepackage{eepic}
\usepackage{makeidx}
\usepackage{epsfig}
\usepackage{url}
\usepackage{supertabular}

\begin{document}
\preprint{http://dx.doi.org/10.1016/j.bpj.2012.12.045, Biophysical Journal}
\bibliographystyle{prsty}

\title{Membrane mediated  aggregation  of curvature inducing nematogens and membrane tubulation}
\author{N. Ramakrishnan }
\email{ram@physics.iitm.ac.in}
\affiliation{Department of Physics, Indian Institute of Technology Madras, Chennai  600036, India}
\author{P. B. Sunil Kumar}
\email{sunil@physics.iitm.ac.in}
\affiliation{Department of Physics, Indian Institute of Technology Madras, Chennai  600036, India}
\author{John H. Ipsen}
\email{ipsen@memphys.sdu.dk}
\affiliation{MEMPHYS- Center for Biomembrane Physics, Department of Physics and Chemistry, \\
University of Southern Denmark, Campusvej 55, DK-5230 Odense M, Denmark}

\date{\today}
\begin{abstract}
 The shapes of cell membranes are largely regulated by  membrane associated, curvature active, proteins.  We use  a numerical model of the membrane with elongated membrane inclusions, recently developed by us,  which posses spontaneous directional curvatures that could be different along and perpendicular to its long axis.   We show that, due to  membrane mediated  interactions these curvature inducing  membrane nematogens can oligomerize spontaneously,  even at low concentrations, and change the local shape of the membrane.  We demonstrate that  for a large group of such inclusions, where the two spontaneous curvatures have equal sign, the tubular conformation and sometime the sheet conformation of the membrane are  the common equilibrium shapes.  We elucidate the factors necessary  for the formation of  these {\it protein lattices}. Furthermore, the elastic properties of the tubes, like their compressional stiffness and persistence length are calculated. Finally, we discuss the possible role of nematic disclination in capping and branching of the tubular membranes. 
\end{abstract}



\maketitle

\section{Introduction}
Membrane shape deformations are key phenomena in a multitude of  cellular processes, including protein sorting, protein transport, organelle biogenesis and signaling. In the last decade  a profusion of regulatory proteins facilitating such shape changes of the cellular membranes has been unravelled, with the BAR protein superfamily  
\cite{Camilli-2009}, the Pex11 family \cite{Opalinska-2011} and coat proteins \cite{Antonny-2006} 
as notable examples. The possibility of such mechanisms has long been anticipated in the biophysical
literature\cite{Sackmann-1984,Leibler-1986}.  However the experimental and theoretical difficulties
involved have hampered the establishment of a quantitative basis for interpreting such phenomena in cell
biology. Recently, we had overcome one such obstacle by the establishment of a  computer simulation technique  
to study how the cooperative effects of membrane inclusions, imposing a curvature along the direction of its orientation, 
remodels vesicular membranes\cite{Ram-2010}.

 In this work we  aim at describing, from a theoretical point of view, the effect of a large group of these membrane curving proteins, 
which can be considered as effectively elongated objects in the plane of the membrane. We consider inclusions  with approximate $\pi$-symmetry,
 i.e. the protein can be considered as essentially indistinguishable  from its form rotated by 180$^\circ$ around the protein center 
in the plane of the membrane.  The membrane inclusions we consider, 
has thus some similarity with nematogens in 3-D nematic liquid crystals. However, they are embedded  in a membrane and may couple 
to its geometry, and it is only the part of the protein in contact with the membrane, which will be subject to these symmetry requirement. 
Therefore, we cannot  consider these membrane inclusions as  
simple liquid crystal nematogens restricted to Euclidean  two dimensional surfaces. In the following we will refer to such membrane inclusions as membrane nematogens. Large groups of membrane curving proteins fall into this category of membrane nematogens.
  An example is  the BAR proteins(\textsf{proteins containing both BAR domains and/or N-terminal helices}), where both the N-terminal 
amphipatic helices and the banana-shaped,  positively charged, dimeric interface with the membrane, induces directional curvature
 \cite{Peter-2004,Weissenhorn-2005,McMohan-2005,Blood-2006,Schulten-2008,Zimmerberg-2009}. 
The caveolin protein family\cite{Williams-2004}, which form dimers and are bound to the membrane by a pair of hairpins and  
the  reticulon,  DP1 and Yop1p involved in the formation of smooth ER \cite{Voeltz-2006,Kozlov-2010}, and are anchored to the membrane 
by two similar hairpins are also examples.  
The cell biology literature has provided good evidence for that the insertion of amphipathic helical peptide sequences not 
only provide a binding mechanism, but also gives rise to local modulation of the membrane curvature \cite{Farsaq-2002, McMahon-2006}. 
More solid, quantitative support for this conjecture is given from biophysical experiments \cite{Helene-2008} and 
theory \cite{Kozlov-2-2008} . Furthermore, biophysical studies  has demonstrated that curvature active membrane inclusions
 have dramatic effects on the cooperative behavior with a complex interplay between lateral ordering and membrane shape.  
However, the detailed mechanisms leading to the specific complex membrane-protein structures have not been characterized. 
This work will elucidate some aspects of these mechanisms for the membrane nematogens. 

Some of the key processes involved in the structural organization of membrane nematogens described in the cell biology literature, 
can be categorized  as follows: (1) the  aggregation of the nematogens - the process where membrane proteins upon
 activation and/or binding to the membrane spontaneously aggregate and form functional cluster of proteins in the membrane\cite{McMohan-2005,Shimada-2007,Blood-2006,Schulten-2008},
 (2) Tubulation of membranes, where the aggregate and the membrane develop tube-like membrane 
structures (e.g. sorting endosomes\cite{Kurten-2001,Mayor-1993} and Mitochondrial outer membrane\cite{Shepard-1999}, 
formation of T-tubules in {\it Drosophila}\cite{Razzaq-2001})  and (3) The formation of {\it  protein lattices}, often characterized by helical 
arrangement of the proteins  spiraling around the tubular membrane, e.g. for dynamin \cite{Sweitzer-1998,Low-2009} 
or caveolin \cite{Post-2010}. 

In this work, we will demonstrate by numerical analysis of a possible physical model, which captures the  membrane conformations 
and the organization of in-plane nematogens, that the above mentioned processes directly results from the cooperative thermodynamic 
behavior of the nematogens coupled to the flexible membrane. Also, we will discuss aspects of the stability of membrane tubes 
and the formation of the edges for membrane sheets. Our model gives a coarse description of the membrane, which  capture properties 
of the membrane which are essential for its large scale organization.  Despite the simplicity of the model, the parameter space is 
too large for a comprehensive discussion of it's phase behavior. Rather, we will  focus on some generic features of the model 
which may well 
give a framework for interpreting the experimental observations of cellular membrane morphogenesis. Previously, 
protein induced membrane tube formation has been considered by a phenomenological model involving 
scalar fields \cite{Sens-2004}, and the coupling between membranes  and inclusions with directional
curvature was modelled in \cite{Fournier-1996,Fournier-1998,Iglic-1999,Fournier-1999}.  

The paper is organized as follows: In Section \ref{sec:model} the physical model of the interacting system of membrane and membrane 
nematogens are presented, while details about the numerical analysis is given in {\bf Supplementary Materials}. 
Section \ref{sec:results} on {\bf Results and Discusion}  present some generic
properties of the model and discuss their possible relevance to experimental results.  
In section \ref{ssec:oligomerize}-\ref{ssec:lattices} the aggregation of proteins and membrane domain formation, 
membrane tubulation and formation 
of  {\it  protein lattices}  are described in the framework of the model. Section \ref{ssec:elasticity} discusses the elastic properties of membrane tubes  and their relevance to observable effects. 
Much of the characterization of the elasticity of {\it  protein lattices} is based on a continuum version of the model 
discussed  {\bf Supplementary Materials}.  
  Section \ref{ssec:endcapping}  discusses mechanisms of closing, capping and branching of  membrane tubules and the possible
 role of nematic point defects.  
Section \ref{activate-segregate}
 describes aspects of the stability of membrane tubules with more membrane curvature  components. 
In section \ref{ssec:sheet-tubes} the interplay between sheet and tubule formation is described and possible implications
for cell organel morphology is discussed.
Some perspective on the modeling of membrane morphogenesis is given in {\bf Conclusion}, Section \ref{sec:conclusion}.  
We will in this work specialize to properties of membranes with inclusions which posses directional curvatures of equal sign. We will consider cases with different signs of the directional curvatures in a seperate publication.
\section{Model} \label{sec:model}
 The modeling of the effects of in-plane nematogens on membrane structure, will in this work, be treated with a discretized description of the surface as a randomly triangulated mesh. A continuous surface conformation is approximated  by a collection of triangles glued together to form a closed surface of well defined topology. A triangulated surface, with spherical topology, thus consist of $N$ vertices connected by $N_L=3 (N-2)$ links, which enclose $N_T=2(N-2)$ triangles. Each vertex $v$ is assigned a position $\vec{X}_v$. This tesselatations of the surface form the basis for a coarse grained description of the membrane, where only the gross features of the structure and interactions are important.
\begin{figure}[!h]
\centering
\includegraphics[width=12cm,clip]{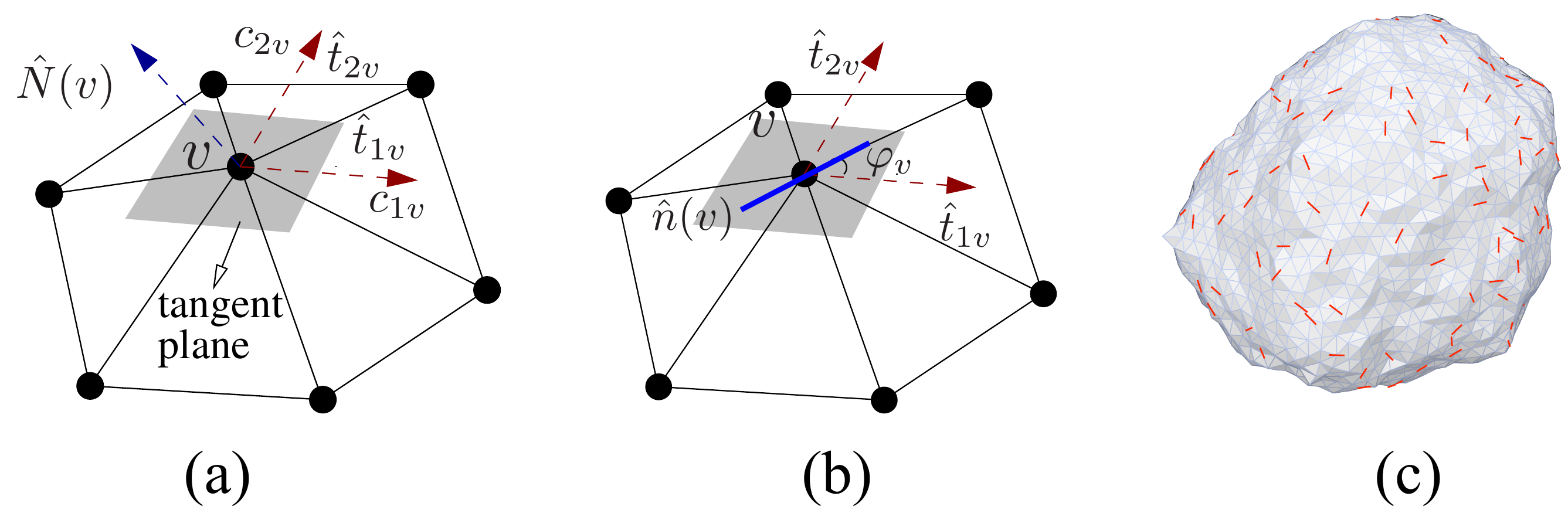}
\caption{ \label{fig:patchmodel} (a) A one ring triangulated patch around a vertex $v$. The shaded region represents the tangent plane at $v$ and $\hat{N}(v)$ its corresponding normal. $c_{1v}$ and $c_{2v}$ are the maximum and minimum principal curvatures, respectively, along principal directions $\hat{t}_{1v}$ and $\hat{t}_{2v}$. (b) Illustration of the nematic field vector $\hat{n}$ defined on the tangent plane of vertex $v$. (c) A vesicle of spherical topology with spatially random  surface nematogens.}
\end{figure}
 
The triangulation and the vertex position form together a discretized surface, a patch of which is given in fig:\ref{fig:surface}. The geometry of the continuous surface, which is approximated by the triangulated surface, can now be characterized by a number of surface quantifiers, e.g. the curvature tensor, the principal directions $(\hat{t}_{1v}\, \&  \,\hat{t}_{2v}$), the corresponding principal curvatures ($c_{1v}$ \& $c_{2v})$ and 
surface normal, $\hat{N}(v)$, at each vertex $v$. The details can be found in \cite{Ram-2010}. 
The discretized Helfrich's free energy\cite{Helfrich-1973} can then be evaluated as  
\begin{equation}
 {\cal H}_c =\frac{\kappa}{2} \sum_{v=1}^N \frac{A_v}{3} H_v^2 
\label{eq:discrhelfric}
\end{equation}

\noindent where $H_v=(c_{1v}+c_{2v})/2$ the mean curvature at vertex $v$
and $A_v$ is the area of the surface patch occupied by the triangles adjacent to vertex $v$. $\kappa$ is the bending
rigidity of the membrane. Furthermore we are in a position to calculate the directional curvatures along  
and perpendicular to  a unit vector $\hat{n}$ along the surface  by use of Gauss formula:

\begin{eqnarray}
H_{v,\parallel} &=& c_{1v} \cos^2 \varphi_{v}+c_{2v} \sin^2 \varphi_{v}, \nonumber\\
H_{v,\perp} &=& c_{1v} \sin^2 \varphi_v+c_{2v} \cos^2 \varphi_{v},
\end{eqnarray}

\noindent where $\varphi_v$ is the angle between $\hat{n}$ and the principal direction $\hat{t}_{1v}$. 
Such an orientational spontaneous curvature may be induced by a membrane nematogen with an orientation in the plane of the membrane given by $\hat{n}$.  In addition to the interaction with the membrane, nematogens may tend to orient along each other at close proximity due to steric, electrostatic and dispersion  interactions  \cite{Onsager}.  In the present study, we focus only on the two dimensional orientational interactions promoted by the underlying, non-planar, fluctuating membrane \cite{Park:1996,Kim:1998,Chou:2001bma,Lewandowski:2008df,Lewandowski:2009cr}.  The $\pi$-symmetry of the individual nematogens dictates that the simplest form of their self interaction should be of the type $\cos^2(\theta_{vu})$ and $\sin^2(\theta_{vu})$, where $\theta_{vu}$ is the angle between $\hat{n}_v$ and $\hat{n}_u$ at neighboring vertices.  We choose to represent the interactions between membrane nematogens by an extension of  the well-established Lebwohl-Lasher model of nematic ordering in presence of vacancies, here implemented on a triangulated surface model of a membrane. 
 
 The nearest neighbor interaction between the nematogens is composed of an isotropic component represented by an
interaction strength $J$ and an anistropic (quadrupolar) correction measured by the interaction constant $\epsilon_{\rm LL}$. The total interaction between the membrane nematogens thus takes the form 
\begin{equation}
 {\cal H}_{\rm field}= \sum_{\langle vu \rangle}\left\{-\frac{J}{2}-\epsilon_{\rm LL}
 \left( \frac{3}{2} \cos^2(\theta_{vu}) -\frac{1}{2}\right)\right\} I_v I_u,
\label{Eq:lebwohl}
\end{equation}
\noindent
where the sum is over nearest neighbour vertices. $I_v=0,1$ is an occupation variable, which is unity if vertex $"v"$ is occupied by a nematogen and zero if otherwise. The calculation of the $\theta_{vu}$ is non-trivial, since the angle between 
spatially separated nematogens are measured after the parallel transport of vectors along the curved surface~\cite{Ram-2010}. With this measure of the angular differences, Eq.(\ref{Eq:lebwohl}) models the in-plane  interactions of the nematogens mediated by the membrane. The direct distance dependent interactions through the cytosol is not taken into account in this model of  membrane-protein conformations. Sufficiently large, positive  $\epsilon_{\rm LL}$ favors in-plane ordering of the nematogens. The effect on the anisotropic elasticity of the membrane due to the nematogens, in analogy with the discretized Helfrich free energy, takes the form~\cite{Ram-2010}: 
\begin{equation}
{\cal H}_{nc}  =  
            \sum_{v=1}^{N} \left \{\frac{\kappa_{\parallel}}{2} \left(H_{v,\parallel}-H_{\parallel}^0\right)^2  \\
             +\frac{\kappa_{\perp}}{2} \left(H_{v,\perp}-H_{\perp}^0\right)^2 \right \} I_v A_v    
\end{equation}

\noindent
$H_{\parallel}^{0}$ and $H_{\perp}^{0}$ are  the spontaneous curvatures along ${\hat n}$ and ${\hat n}^{\perp}$, while
$\kappa_{\parallel}$ and $\kappa_{\perp}$ are the corresponding directional membrane bending elastic constants.

\noindent
Self avoidance of the discretized surface is ensured by imposing constraints on the neighboring vertex distance and
on the dihedral angles between neighboring faces\cite{Ram-2010}. The equilibrium properties of the discretized surface can now be evaluated by standard Monte Carlo sampling of Boltzmann's probability distribution 
$\sim \exp\left(-\frac{1}{k_B T} [{\cal H}_c+ {\cal H}_{\rm field}+{\cal H}_{nc}]\right)$ at fixed concentrations of 
the membrane nematogens.  A general description of the implemetation of such numerical models and 
further details about the simulations are given in \cite{Ram-2010}. 

Finally, we will make some considerations about length scales. The lattice model is a highly coarse-grained representation of  the membrane,  designed to capture the large length-scale properties of membranes with inclusions. Therefore, the triangulated surface represent a collection of membrane patches with a characteristic length scale. A natural choice of length scale is to identify a tether length with the lateral extension of a membrane inclusion. Some examples here are CIP4 F-BAR with a length of 22 nm\cite{Camilli-2008} or dynamin which extent 
about 25 nm \cite{Sweitzer-1998}. 

\noindent
The computer simulations of the discrete model provide us with insight into the nature of equilibrium configurations for a choice of model parameters. To complement the numerical simulations, it is useful to consider the corresponding continuum model in the limit of membrane nematogen with 100\% surface coverage. It is an extension of Helfrich's 
bending free energy functional \cite{Peliti-1989,Frank-2008}

\begin{eqnarray}
\label{eqn:totham}
{\cal F}& =& \oint dA \left\{ \frac{K_A}{2} {\rm Tr}\left(\nabla\hat{n}:\nabla\hat{n}\right) + \frac{\kappa}{2} (2 H)^2 \right. \nonumber\label{Eq:nematic}\\
            && \left. + \frac{\kappa_{\parallel}}{2}(H_{n,\parallel}-H^{0}_{\parallel})^2 +
              \frac{\kappa_{\perp}}{2}(H_{n,\perp}-H^{0}_{\perp})^2\right\} \nonumber  
\end{eqnarray}

\noindent
where $K_A=\frac{3\sqrt{3}}{2}\epsilon_{LL}$.
In {\bf Supplementary Material} is presented  such an analysis of the mechanical properties of a tubular membrane with a protein coat and analytical expressions reflecting, tube radius, persistence length and protein organization are also derived.

\section{Results and Discussion} \label{sec:results}
 In this section we will present some key aspects resulting from the coupling of membrane nematogen  proteins to lipid membranes.  It will both contain results from computer simulations of the aforementioned model, which are non-perturbative, along with theoretical analysis of the continuum model, of  a more perturbative character to  qualify  the numerical finding. Throughout the discussion the parameter $\epsilon_{LL}$ has a relatively high value (several $k_B T$ in a range where nematic ordering is favored). Furthermore, the  implications of our results on the experimental systems {\it in vivo} and {\it in vitro} will be discussed.  
 
 \subsection{Aggregation and membrane domain formation of membrane nematogens} \label{ssec:oligomerize}
\begin{figure}[!h]
\centering
\includegraphics[width=12.5cm]{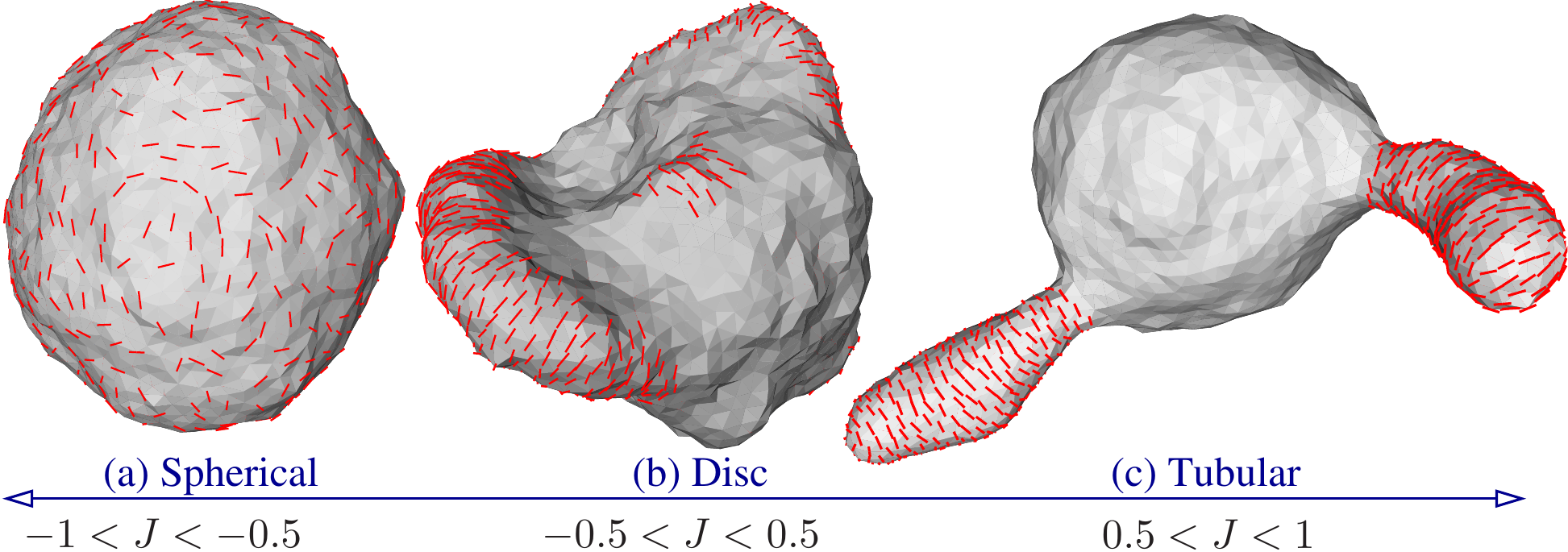}
\caption{\label{fig:conffuncJ} Equilibrium membrane conformations, with $\phi_{A}=0.3$, $\kappa=20$, $\kappa_{\parallel}=5$ and $H^{0}_{\parallel}=0.5$ and $\epsilon_{LL}=3$, for different range of $J$.}
\end{figure}

 A common feature of membrane nematogens  is their strong tendency to self-associate, driven by the  
  flexible  geometry of the membrane - in this manuscript, we call this self associated structure to be an aggregate or a domain. Self association has been observed for a wide range of model parameters $\kappa_{\parallel}$, $\kappa_{\perp}$, $H_{\parallel}^0$ and $H_{\perp}^0$. 
  All results presented in the following corresponds to system size with $N=2030$ vertices.  
  When the fraction  of nematogens $\phi_{A}=0.3$, $\epsilon_{LL}=3$ and  $J=0$ (in units of $k_B T$), 
 the flexible membrane with  curvature coupled to the nematic  orientation,  gives rise to co-existence  
of nematically ordered domains and the isotropic dilute phase,  this is shown in figure-2(b). 
 This is to be compared with the  planar Lebwohl-Lasher model on a random triangular lattice,  at the same concentration, where the  isotropic phase is stable  ({\sf see Supplementary Materials}). 
Additional direct repulsive interactions $J\leq-0.5$ between the membrane nematogens can reestablish the isotropc phase, which is shown in  Fig. \ref{fig:conffuncJ}(a).  The  aggregation  of membrane nematogens  cause shape deformation of the whole membrane  with the  collective involvement  of all the  degrees of freedoms,  lateral orientation, lateral position and membrane conformation. In general the  lateral domain formation depends on all the involved parameters - e.g. increasing $J$ promotes the  aggregation and can change the aggregate shape as shown in Fig. (\ref{fig:conffuncJ})(c).

\begin{figure}[!h]
\begin{center}
\includegraphics[width=15cm]{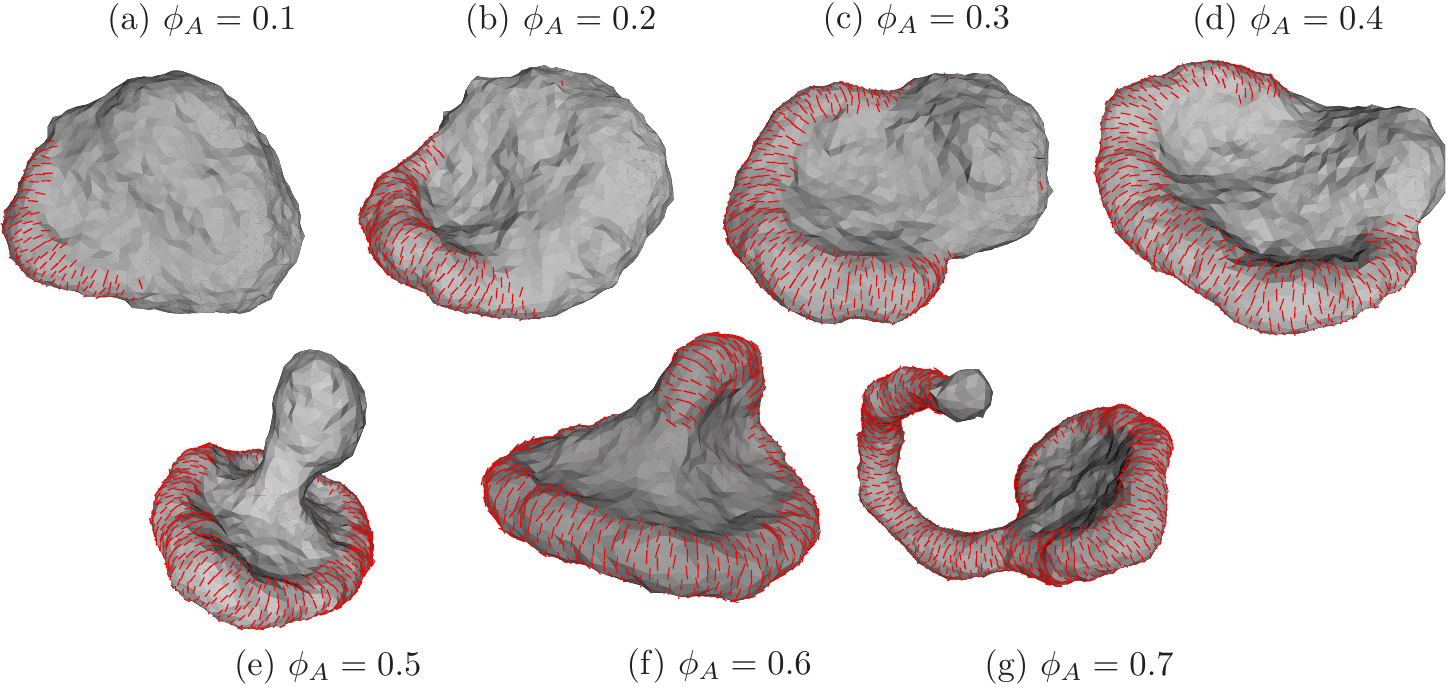}
\end{center}
\caption{Equilibrium configurations for varying composition. $\kappa=10$, $\kappa_{\parallel}=5$, $\kappa_{\perp}=0$, 
 $H_{\parallel}^{0}=0.5$, $H_{\perp}^{0}=0$, $J=0$, $\phi_{A}=0.1- 0.7$ and $\epsilon_{LL}=3$.}
\label{fig:czero0p5-funcconc}
\end{figure}

\noindent
The effect of concentration is shown in Fig.(\ref{fig:czero0p5-funcconc}) for surface coverage in the range  $\phi_{A}=0.1\,-\,0.7$, which display a series  of complex shape deformations connected to different aggregate structures. More details will be discussed in section \ref{ssec:sheet-tubes}.

\begin{figure}[!h]
\begin{center}
\includegraphics[width=12.5cm]{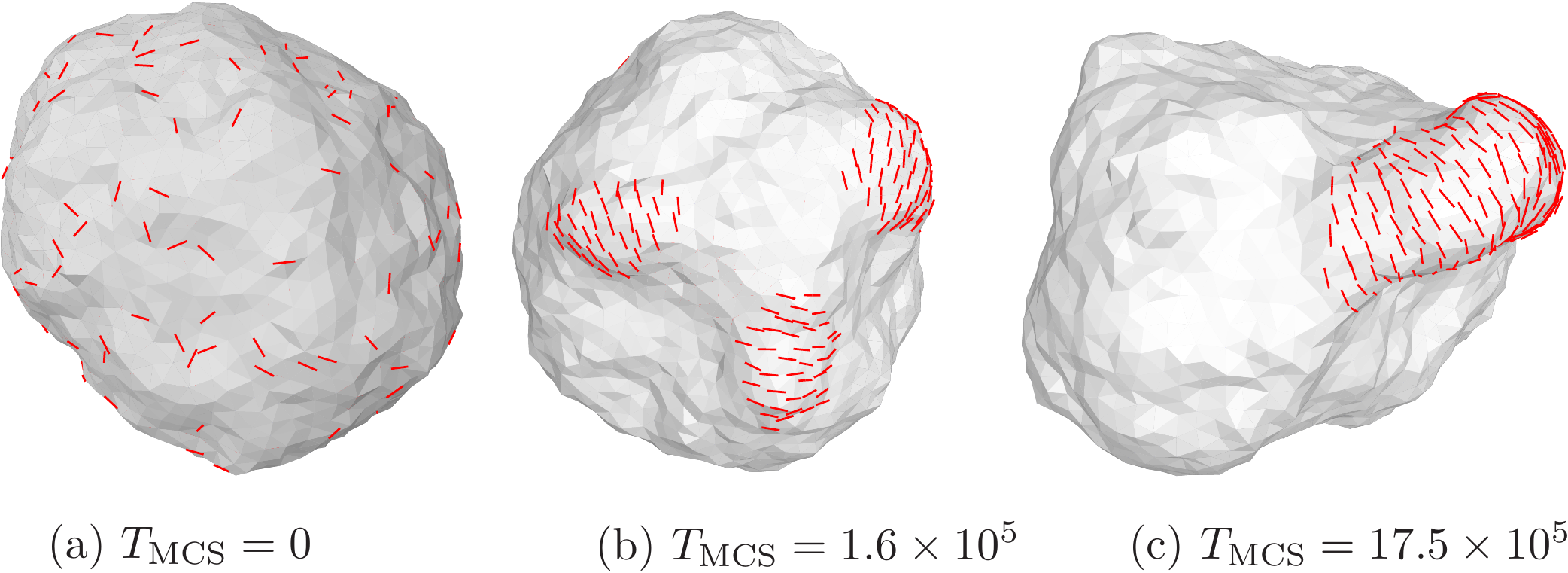}
\end{center}
\caption{ Aggregation of membrane inclusions for $\kappa=10$, $\kappa_{\parallel}=5$, $\kappa_{\perp}=0$, 
 $H_{\parallel}^{0}=0.5$, $H_{\perp}^{0}=0$, $J=0$, $\phi_{A}=0.1$ and $\epsilon_{LL}=3$.  Monte Carlo time series showing a) random initial configuration of membrane nematogens, b) intermediate state with multiple nematic domains and c) equilibrium conformation where all the small domains coarsen into a single patch.} \label{fig:oligomerize-10per}
\end{figure}

The  aggregation of membrane nematogens also has a temporal aspect. In Fig.(\ref{fig:oligomerize-10per})  we have shown a Monte Carlo time series, for a membrane coverage of 10\% nematogens, to illustrate the sequence of   domain formation and membrane curvature  induced changes leading to the equilibrium structure. The membrane nematogens in an initial randomly dispersed orientation assemble into smaller orientationally  
ordered domains mediating the final equilibrium structure. These  ordered domains often appear as metastable configurations, which either disperse again due to lateral fluctuations or they will eventually funnel into a  equilibrium domain configuration.  The Monte Carlo dynamics does not reflect the physical kinetics very well, but is useful in identifying kinetic paths connecting various metastable states  that lead to the global minima \cite{Sunil-1998,Sunil-2001} . 

\noindent
The   aggregation  of membrane inclusions mediated by membrane curvature deformations and fluctuations is not 
specific for nematogens, but is a more general phenomena for membrane curvature active components. 
It has been well understood in the framework of models for curvature instabilities \cite{Leibler-1986,Helene-2008,Henriksen-2010}, 
and has also been demonstrated that simple amphiphathic inclusions, e.g. antimicrobial peptides like 
Magainin or Melittin \cite{Helene-2008,Gerbeaud-1998} and viral membrane active proteins  like NSB4 
of Hepetites C\cite{Penin-2010}. \\

The self-association of these membrane components thus needs not to be facilitated by strong direct attractive interaction amongst them. The coupling to the membrane geometry provides  additional indirect membrane conformation mediated attractive forces making them to slip into bound structures  involving both the proteins and membrane curvature. However, the structure of the aggregates are dependent on the details of the molecular structure and the direct interactions. A general feature of these   aggregates is that they appear as nematically ordered domains, where the nematogens form  elongated oriented patches with well defined curvature characteristics, e.g. ridges or cylindrical rims. In the following we will in particular focus on the tube-like structures. 

 \subsection{Tube formation} \label{ssec:tubulation}
 The most prevalent equilibrium domain structure is the nematic tube, where the membrane protrude into a cylinder like configuration with the membrane nematogens forming a coat around the cylinder.  Also for tube formation the overall interaction strength ($J$) between the membrane nematogens plays a secondary role. It's most pronounced effect is to widen the concentration range for tubulation and to enhance the line tension at the domain boundary, which can induce fission of tubes by narrowing
the tube at the  boundary of the domain, as in Fig(\ref{fig:conffuncJ}(c)).  The effect of concentration of membrane nematogens on the membrane tubulation phenomena is shown in Fig.(\ref{fig:czero0p5-funcconc}(g)). For large concentrations of nematogens or increasing values of $J$ the tubes are the characteristic equilibrium structures shown in Figs. (\ref{fig:conffuncJ},\ref{fig:czero0p5-funcconc}). 

The radius of the equilibrium membrane tubes appears to be relatively well-defined. The radius of the tube with nematic order can be calculated on basis of the continuum model Eq.\eqref{eqn:totham} for the chosen model parameters (see {\bf Supplementary Material}):

\begin{equation}
\bar{r} =\left\{ \begin{array}{ll}
    \frac{1}{|H_{0 \parallel}|}\sqrt{\frac{\kappa_{\parallel} + \kappa}{\kappa_{\parallel}}} & {\rm for}\ \ \kappa_{\perp}=0 \\
    \frac{1}{|H_{0 \perp}|}\sqrt{\frac{\kappa_{\perp} + \kappa}{\kappa_{\perp}}} & {\rm for}\ \ \kappa_{\parallel}=0  
    \end{array} \right.
\label{eq:delta}
\end{equation}

\noindent
So, the radius $\bar{r}$ is set by the curvature elastic model parameters, involving the absolute value of the directional 
spontanous curvatures, modulated by the curvature elastic constants. It follows from Eq.(\ref{eq:delta}) that the actual tube 
radius is somewhat larger than the inverse directional spontaneous curvatures and dependent on the relative strength of the elastic constants.

In experimental systems the membrane tube dimensions  can vary considerably with different types of proteins in the 
cell\cite{Camilli-2009}.  Membrane tubes formed {\it in vitro} by curvature active proteins also display a considerable 
variability  in size. Frost et al. \cite{Camilli-2008}  have studied the effect of a number of mutants of  CIP4 F-BAR on liposomes.  
By mutations they find a big variations of  tube diameters in the range of 50 to 100 nm.

\begin{figure}[!h]
\centering
\includegraphics[width=15cm]{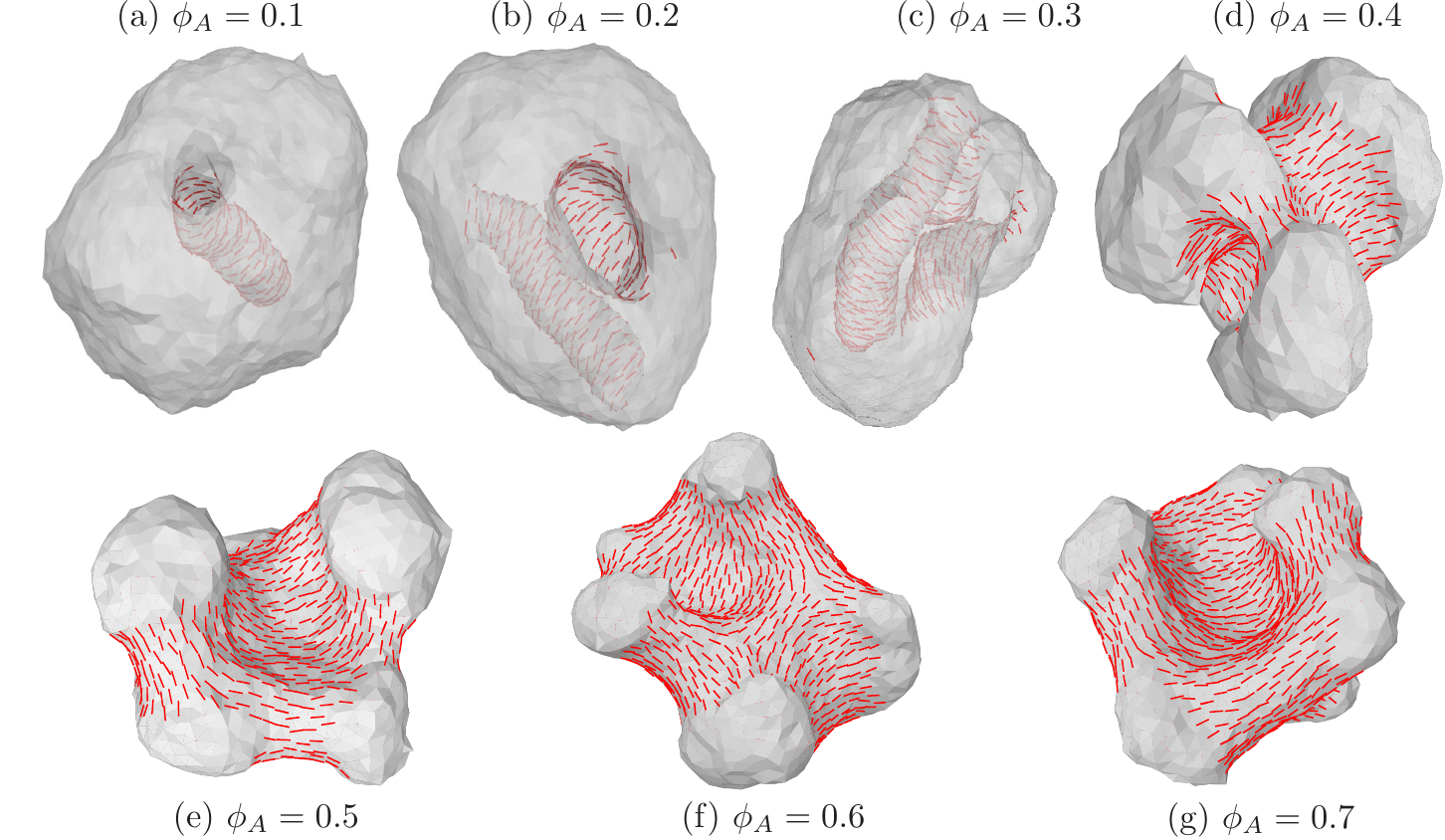}
\caption{Equilibrium configurations for vesicle with negative spontanous curvatures. $\kappa=10$, $\kappa_{\parallel}=5$, $\kappa_{\perp}=0$, $H_{\parallel}^{0}=-0.5$, $H_{\perp}^{0}=0.0$, $J=0$ and $\phi_{A}=0.1\,-\,0.7$.  \label{fig:czerom0p5-funcconc} }
\end{figure}

Membrane tubes induced by membrane inclusions are common phenomena in biological cells, both as more static structures like T-tubules of the muscle cells\cite{Razzaq-2001} or more temporal structures like sorting endosomes \cite{Hurley-2008}. The examples shown in Fig.(\ref{fig:conffuncJ},\ref{fig:czero0p5-funcconc}) corresponds to the
 cases where spontaneous curvatures are positive, like  that induced  when F- BAR-domain proteins bind to organelle membranes.
 However, if  the proteins induce negative spontanous curvatures, as in I-BAR domain proteins,  it gives rise to tubular invaginations  as shown in Fig.(\ref{fig:czerom0p5-funcconc}) .

As can be seen from Fig.(\ref{fig:czerom0p5-funcconc}), for proteins with large negative spontaneous directional curvatures,  at low concentration ($\phi_{A}=0.1-0.3$), we obtain  tubes growing into the interior of the vesicle.  As $\phi_{A}$ increases,  tubes disappear and saddle like regions  appear.  The inner tubes and saddle like regions coexist again  for large concentrations  $\phi_{A}>0.8$.

 \subsection{Protein lattices} \label{ssec:lattices}
\begin{figure}[!h]
\begin{center}
\includegraphics[width=12.5cm]{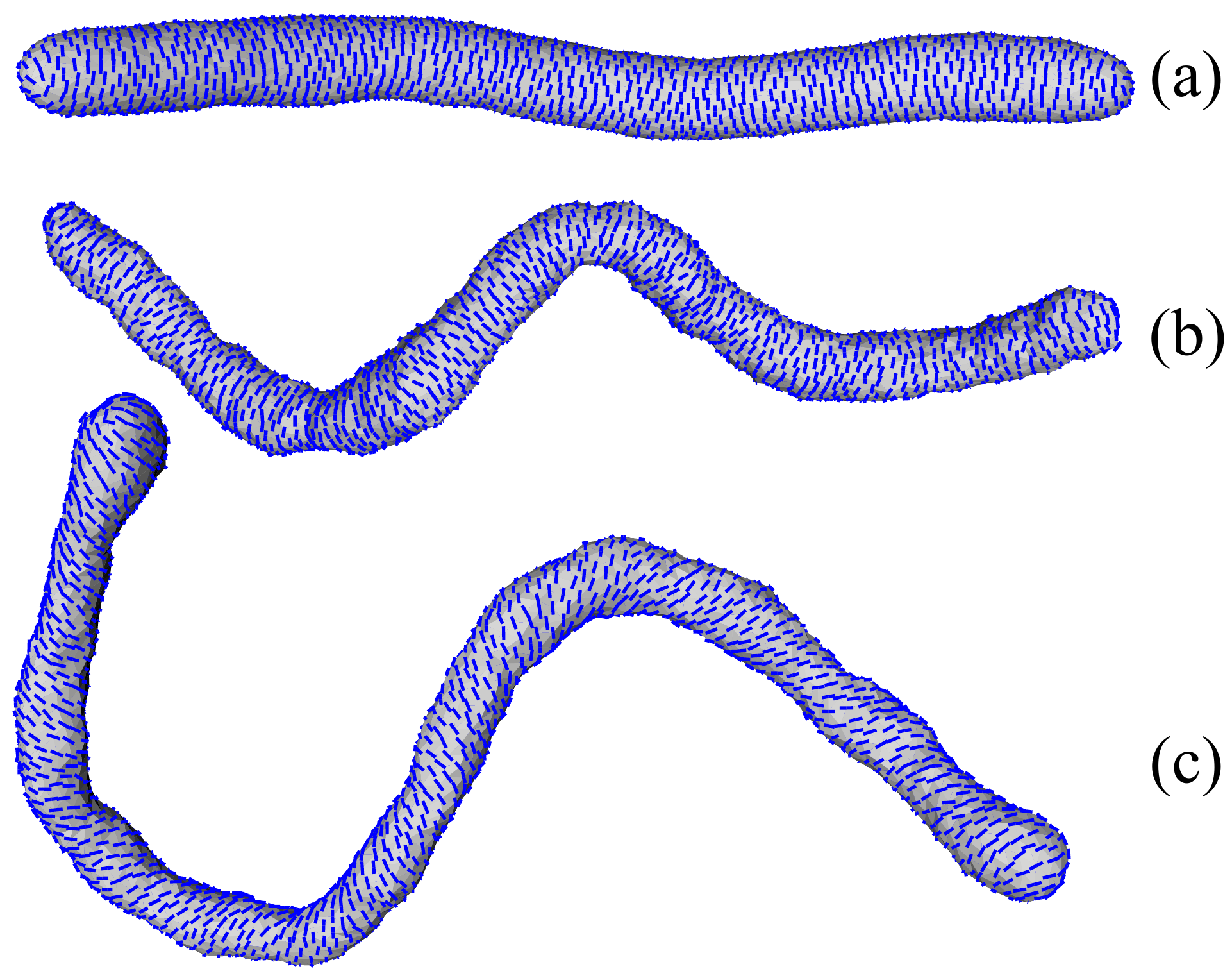} 
\end{center}
\caption{ \label{protein_lattice} {\it Protein lattices}.  Modes of a tubular membrane at different state points, for $\kappa=10$ and $\epsilon_{LL}=3$. {\textbf (a)} Tubular conformation with $\langle \varphi \rangle = 0$ for $\kappa_{\parallel}=5$, $\kappa_{\perp}=0$, $H_{\parallel}^{0}=0.4$, {\textbf (b)} Spiral modes of the tube with $\langle \varphi \rangle = 0$ seen for $\kappa_{\parallel}=5$, $\kappa_{\perp}=0$, $H_{\parallel}^{0}=0.4$,  and {\textbf (c)} rearrangement of nematics into spiral modes ($\langle \varphi \rangle \neq 0$) when $\kappa_{\parallel}=5$, $\kappa_{\perp}=5$, $H_{\parallel}^{0}=0.4$, $H_{\perp}^{0}=0.25$ }

\end{figure}

 Membrane nematogens organize as nematically ordered domains and coat around the membrane to form  tubes. Nematogens orient perpendicular to the tube axis when $\kappa_{\perp}=0, \kappa_{\parallel}\neq 0$ and $H_{\parallel}^{0}>0$.  Similarly, $\kappa_{\parallel}= 0$, $\kappa_{\perp}\neq 0$ and  $H_{\perp}^0\neq0$ leads to an arrangement of the nematogens along the tube direction.  For the common membrane nematogen both these parameters are non-vanishing. Such a case is shown in Fig.(\ref{protein_lattice}).

 The helical arrangement of the membrane nematogens at the tube surface can be easily understood considering that in general 
such arrangement will give rise to a global nematic ordering of the membrane nematogens (generalized spirals are the only 
geodesic curves on long tubes) and the radius is set by the elastic terms. The coupled expressions for the mean values for  
tube radius ${\bar r}$ and the angle ${\bar \varphi}$ between the tube direction   and the nematogen orientation, for different 
regimes of the  dimension less parameter $\tilde{\psi}$:

\begin{equation}
\bar{r} =\left\{ \begin{array}{ll}
    \sqrt{\dfrac{\kappa_{\perp}+\kappa}{\kappa_{\parallel}(H^{0}_{\parallel})^2+\kappa_{\perp}(H^{0}_{\perp})^2}}
    & {\rm for}\ \ \ \tilde{\psi}\leq 0,\\
    &   \\
 \sqrt{\dfrac{\left(\frac{\kappa}{2}\right) \left(\kappa_{\parallel}+\kappa_{\perp}\right) + \kappa_{\perp} \kappa_{\parallel}}{\kappa_{\perp}\kappa_{\parallel}\left(H_{\parallel}^{0}+H_{\perp}^{0}\right)^2}}
   & 0<\tilde{\psi}<1 \\
   &     \\
\sqrt{\dfrac{\kappa_{\parallel} + \kappa}{\kappa_{\parallel}(H_{\parallel}^{0})^2+\kappa_{\perp}(H_{\perp}^{0})^2}}
 & \tilde{\psi}\geq 1
    \end{array} \right.
\label{eq:expforr}
\end{equation}

\noindent
The parameter $\tilde{\psi}$ is given by the model parameters as:
\begin{equation}
\tilde{\psi}=\dfrac{\kappa_{\parallel}H_{\parallel}^{0}-\kappa_{\perp}H_{\perp}^{0} }
{|H_{\parallel}^{0}+H^{0}_{\perp}|(\kappa_{\parallel} + \kappa_{\perp)}}
\sqrt{1+\frac{\kappa}{2}\left(\frac{1}{\kappa_{\parallel}}+\frac{1}{\kappa_{\perp}} \right)} + \frac{\kappa_{\perp}}{\kappa_{\parallel}}
\end{equation}

\noindent
similarly for the angle $\bar{\varphi}$:

\begin{equation}
\cos^2(\bar{\varphi})  =\left\{ \begin{array}{ll}
    0 & {\rm for}\ \ \ \tilde{\psi}\leq0,\\
     &   \\
    \dfrac{\left(\kappa_{\parallel}H^{0}_{\parallel}-\kappa_{\perp}H^{0}_{\perp}\right)\bar{r}+\kappa_{\perp}}
{\kappa_{\perp}+\kappa_{\parallel}} & 0<\tilde{\psi}<1 \\
     &   \\
     1 & \tilde{\psi}\geq 1 
    \end{array} \right.
\label{eq:expforx}
\end{equation}

\noindent So, in general both $\bar{\varphi}$ and $\bar{r}$ are set by the model parameters.  A derivation of the above expressions  are given in {\bf Supplementary Materials} .

The spiral organization of the membrane coating proteins has now been observed for many tubular membrabe 
systems {\it in vivo}  and {\it in vitro}, e.g for the F-BAR proteins\cite{Camilli-2008, Low-2009}. 
EM-tomographs of tubules of CIP4 F-BAR on liposomes \cite{Camilli-2008} show a fairly dense packing arrangement in the helical tube. The average tube diameter is around 68 nm and the helical angle is about {\bf $\varphi=40^{\circ}$}.  Such arrangements are termed  {\it  protein lattices} in the cell biology literature. It is found that the helical angle $\varphi$  of the {\it  protein  lattice} with respect to the tube direction adjusts to the tube diameter such that   the directional curvature is about the same. For a similar type of experiment with dynamin\cite{Sweitzer:1998p3632} the membrane tubes of radius $r \simeq 23$nm  with densely packed helical dynamin coat was observed with a helical pitch of about $15$nm  (corresponding to a helical angle $\varphi= 80^{\circ}$).

Our simulation results suggests that the spiral organization of the protein coat on the tube  need not be  a result of
 polymerization as often referred in the literature, but can be  a self-assembly process of the curvature active 
proteins mediated by the membrane.  Furthermore, the modelling suggest that these {\it  protein lattices} are
not  conventional two dimensional  lattice structures like polymerized membranes or graphene, but 
rather two dimensional  nematic liquid crystalline structures. In the model there is no terms which can distinguish between a right or left turning helix, i.e. the helical arrangement is the result of a spontaneous symmetry breaking. 
However, the smallest chiral symmetry breaking contribution to the free energy  can favor one of the helical orientations without having an effect on any other parameters. 

\subsection{Thermal stability of membrane tubes} \label{ssec:elasticity} 
 While our model parameter  determine the   mean physical properties  of the tubes , e.g. the radius,  we expect the tubular membranes 
to display an elastic response to deformations in its shape and   organization of the membrane nematogens. This can, for e.g.,
be reflected in the variation of the shape characteristics due to thermal fluctuations. 
For analysis of such deformations the continuum description  of coated membrane tubes are suitable and the details can be found in
{\bf Supplementary Material}.  
It is shown that in general the deviations in the orientation of the membrane nematogen  and the tube radius are strongly correlated. 
The  thermally induced fluctuations  in the radius is found to be
\begin{equation}
  \frac{\langle (\delta r)^2 \rangle}{\bar{r}^2}
=\frac{k_B T}{4 \pi} \frac{\kappa_{\parallel}+\kappa_{\perp}}{\kappa_{\parallel}\kappa_{\perp}+(\kappa_{\parallel}+\kappa_{\perp})\frac{\kappa}{2}}\ \  {\rm for} \ \ 0<\bar{\psi}<1,
\label{eq:rratio}
\end{equation}

\noindent
where $\bar{r}$ and $\bar{\varphi}$ are respectively the equilibrium tube radius and nematic orientations and  $\bar{\psi}=\cos^{2}\bar{\varphi}$. We note that the relative variance in $r$ has an upper limit $\frac{k_B T}{2 \pi \kappa}$. With a typical range 
$\kappa \sim 20-50{\rm k_B T}$ this ratio in Eq.(\ref{eq:rratio}) is of the order 0.01. 
For CIP4 F-BAR,  reconstituted on liposomes,  cryo-tomography  measurements  give $\bar{r}=33$nm and 
$\langle (\delta r)^2 \rangle/\bar{r}^2\simeq 0.01$ \cite{Camilli-2008}.  
If this observed variation in tube thickness is interpreted as frozen in thermal variations, 
it  is in agreement with the the above theory. 
For rigid membranes with large $\kappa$ and/or large $\kappa_{\parallel}, \kappa_{\perp}$ values we can consider the 
thermaly excited variations in $r$ as small. Similarly, we can estimate the thermal fluctuations around 
$\cos^2(\bar{\varphi})$ for such a segment as, 

\begin{equation}
 \langle (\delta \cos^2(\varphi))^2 \rangle = \frac{k_B T}{4\pi} \frac{\kappa_{\parallel}\psi^2 + \kappa_{\parallel} (1-\psi)^2 + \frac{\kappa}{2}}{\kappa_{\parallel}\kappa_{\perp} + (\kappa_{\parallel}+\kappa_{\perp})\frac{\kappa}{2}}\ \  {\rm for} \ \ 0<\bar{\psi}<1.
\end{equation}

\noindent
To our knowledge no experimental reports on the random variations in the helical angle has been given.
\noindent
A third type of deformation to consider is the bending of the tubes. It is  shown in {\bf Supplementary Materials} 
that when $r$ is a  constant along the tube, the free energy expression is relatively simple. 
In particular we find that the free energy of bending for a tubular membrane takes the approximate form,
\begin{equation}
\Delta F_{\rm tot} \approx \frac{1}{2} k_B T\, l_P \, \int_0^L ds \lambda(s)^2,
\end{equation}

\noindent
where $s$ is the the arc length and $\lambda(s)$  is the line curvature along the tube, while $l_P$ is the persistence length of the tube:

\begin{equation}
 l_P = \frac{ \pi \bar{r}\left(K_A + \kappa + \kappa_{\parallel}(1-\bar{\psi})^2
+\kappa_{\perp} \bar{\psi}^2\right)}{k_B T}.
\label{eq:lp}
\end{equation}

\noindent
There are few experimental measurements of the persistence length of membrane tubes with {\it  protein lattices}. 
 For the F-BAR FBP17 producing tubes of radius $r(FBP17)=34{\rm nm}$ the persistence length  was measured 
to $l_P(FBP17)=142 \mu {\rm m}$ \cite{Camilli-2008} while for amphiphysin $r{\rm (amph)}\sim 7$nm\cite{Roux-2012}  and $l_P{\rm(amph)}=9\mu{\rm m}$  
while for dynamin $r{\rm (dynamin)}\simeq 20{\rm nm} $ and $l_P{\rm (dynamin)}=37 \mu {\rm m}$. 
A calculations of $l_P$ from  Eq.(\ref{eq:lp}) solely based on $\kappa$ gives predictions which are an order of magnitude too small, 
which indicates that other elastic constants $\kappa_{\parallel},\kappa_{\perp}$ and $K_A$ gives the main contributions to $l_P$.

\subsection{Capping the tubes, Defects}\label{ssec:endcapping}
\begin{figure}[!h]
\centering
\includegraphics[height=2in]{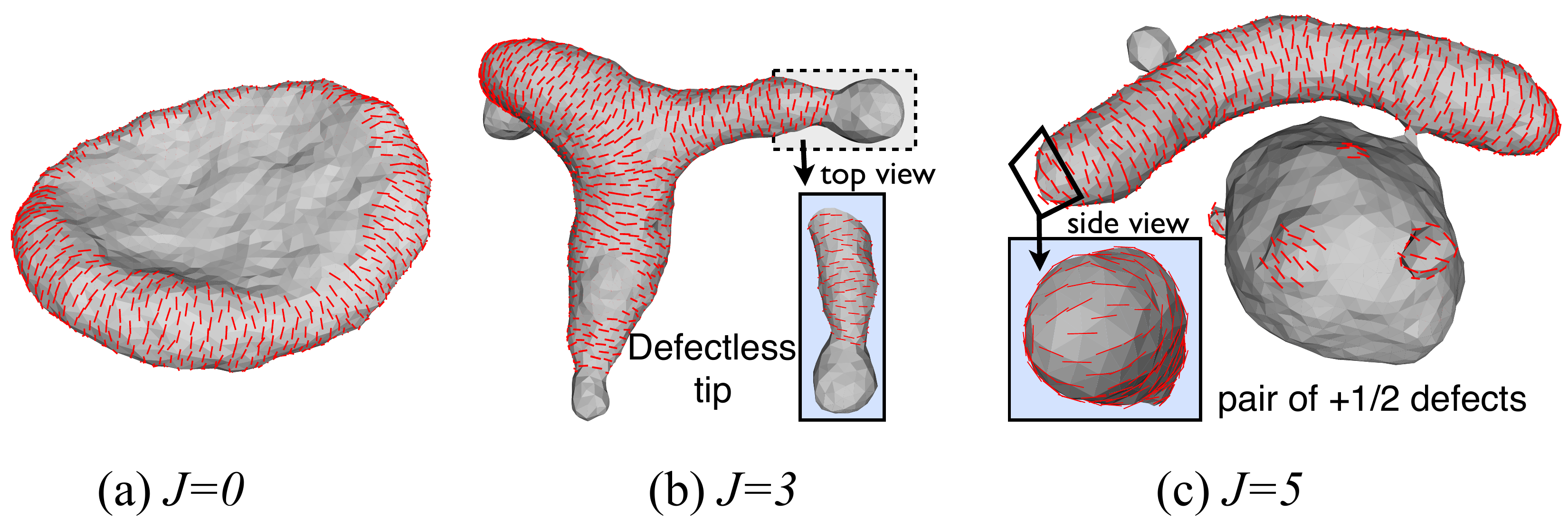}
\caption{\label{fig:comp-defectjzero} A partly decorated membrane with $\kappa=20$, $\kappa_{\parallel}=5$, $\epsilon_{\rm LL}=3$.
 Shown are: (a) a disc without a defect for $\phi_{A}=0.4$, $H^{0}_{\parallel}=0.5$ and $J=0$. (b) Tubes without  defects 
at  $J=3$. ( the  bottom panel shows an enlarged side view of the tip of a tube without defects ).  (c) Tubes and buds  with   
defects when  $J=5$. ( the  bottom panel shows an enlarged top view of the tip of a bud with  defects ) }
\end{figure}

 The formation of membrane tubes with helical coats seems to be generic for systems with membranes with
 membrane nematogens. Either the helical coat  has to terminate  resulting in an  interfacial curve 
separating the coated and uncoated regions or  the vesicle should  sprout tubes and buds with the tips having a  pair of  point defects.
The way this takes place in the tube end or at a domain boundary is mainly determined by the competition between   interfacial tension, 
which in our model is largely regulated by the parameter $J$, and bending modulus. In Fig(\ref{fig:comp-defectjzero}) is shown 
that when the interaction parameter $J$ is increased
the  interfacial line shrinks, first transforming the vesicle from a disk to a  structure with partially coated  tubes and buds, 
but still no defects.  Further increase in the line tension will result in tubes and buds that are fully coated but minimizing the 
length of the interfacial line between coated and uncoated regions.  It does so by either moving  the interfaces  to the end of 
the tube forming  a pair of point defects or  deforming the  membrane to form a narrow neck.  Note that the line does not shrink 
to a single point defect of strength $+1$ but instead forms a pair of $+1/2$ defects bound to each other. This is the result of 
of the $\pi$ symmetry of the membrane nematogen and the strong coupling between membrane curvature and nematic 
orientation \cite{Prost-1992,Vitelli-2004,Ram-2011}. To our knowledge no details about the capping of the coated membrane tubes have been
provided by experiments. 

\subsection{Curvature differences leads to segregation}\label{activate-segregate}

\begin{figure}[!h]
\includegraphics[width=12.5cm,clip]{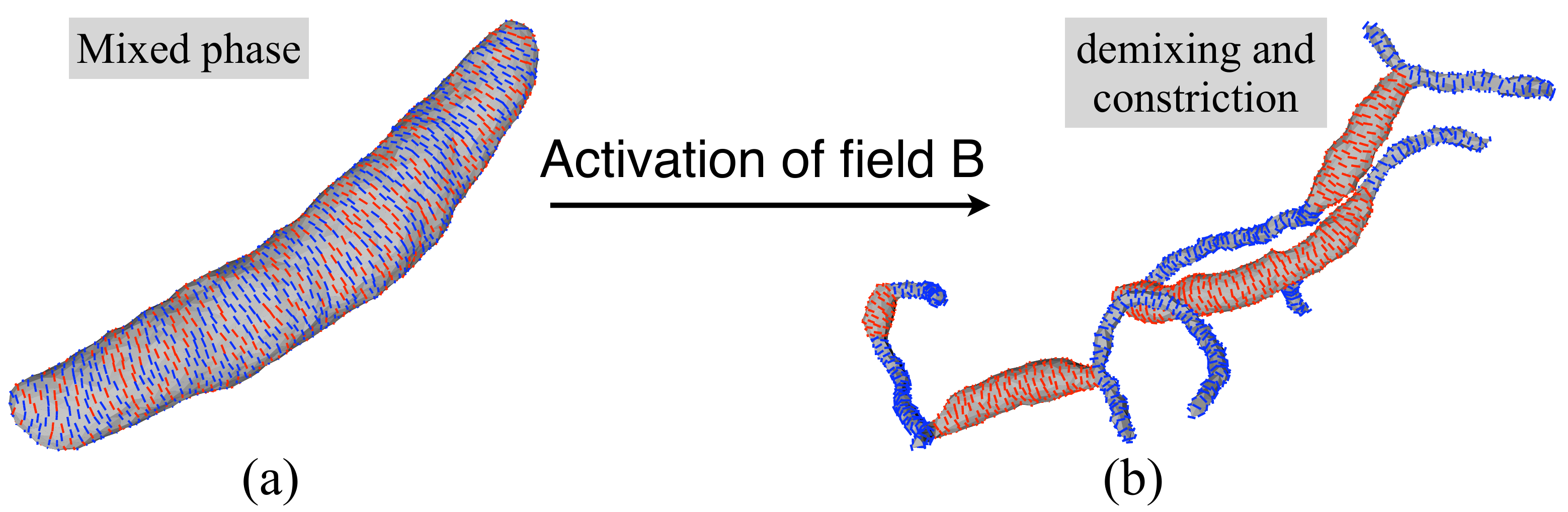}
\caption{\label{fig:uncons-cons} (a) Tubular membrane of uniform cross section with fields A and B in the mixed phase  for $H^{0}_{\parallel,A},H^{0}_{\parallel,B}=0.5,0.5$. (b) Activation of field B takes $H^{0}_{\parallel,B}$ from 0.5 to 1.25. The difference in the spontaneous curvatures leads to phase segregation and results in the constriction of tubes. Field concentrations are respectively $\phi_{A}=0.4$ and $\phi_{B}=0.6$ while $\kappa=10k_{B}T$, $\kappa_{\parallel}^{A}=\kappa_{\parallel}^{B}=5k_{B}T$ and $\epsilon_{AA}=\epsilon_{BB}=3k_{B}T$.}
\end{figure}
Another example of the curvature driven aggregation is demonstrated in Fig. \ref{fig:uncons-cons}. 
Shown in Fig. \ref{fig:uncons-cons}(a) is a  tubular membrane of uniform cross section,  fully decorated  
by two different  types of membrane nematogens, labelled A and B.  The tube is stable, in the mixed state,  when 
the directional spontaneous curvature of the in-plane fields,  $H^{0}_{\parallel,A}=H^{0}_{\parallel,B}=0.5$, are the same.  
If a source of activation  increases the spontaneous directional curvature  of B to $H{0}_{\parallel,B}=1.25$, 
analogous to activation of dynamin proteins by hydrolysis of GTP,  the fields demix.  The regions containing field B constrict  the tube further. 
The equilibrium shape of the activated membrane is observed to have successive tubular regions of large and small 
curvatures ( see  Fig. \ref{fig:uncons-cons}(b)), similar to the striated patterns of dynamin tubes obtained on 
treatment with GTP$\gamma$s~\cite{Roux-2010,Hinshaw-2004}. For dynamin the molecular conformation and membrane tube diameter is GTP dependent \cite{Hinshaw-2001,Hinshaw-2004}. Furthermore it is observed that the tube constriction involves a tube twisting, suggesting a change in the helical angle $\varphi$ \cite{Roux-2006}.  
In {\it in vivo} experiments, it has been demonstrated that structurally similar F-BAR proteins can co-localize into
the same membrane tubes \cite{Shimada-2007} while differing BAR proteins, like F-BAR and N-BAR, seggregate into 
separate membrane tubes with their respective characteristic $r$ and $\varphi$ \cite{Itoh-2006,Camilli-2008}.
Our analysis suggests that this recruitement of differing BAR-proteins into separate domains is possibly driven
by their directional spontaneous curvatures.

 \subsection{Sheets versus tubes}\label{ssec:sheet-tubes}
The effect of concentration, shown in Fig.(\ref{fig:czero0p5-funcconc}), for surface coverages in the range 10\% and 70\%  display a series of complex shape deformations connected to different aggregate structures. 
The regime, where inclusions stay separate,  for the model parameters chosen here,  appear at very low concentrations. 
The figure illustrates that for a system where the direct interactions parameter $J$ between
the membrane nematogens are weak the oligomers tend form larger rim-like formations, which stabilizes disc like structures of vesicles. The rims form the edges of the discs. As the concentration is increased part of the edge turn tubular.  

So for a range of concentrations the disc and the tubes coexist.  The tubules get more pronounced and the discs diminishes  with 
the  increase in concentration of membrane nematogens. Recent experiments on the formation of tubular (or smooth) ER suggest that
 some  membrane curvature active proteins, reticulon protein and DP1\cite{Voeltz-2006}, are highly enriched in 
the tubular ER \cite{Shibata-2008} and the ER sheet edges\cite{Kozlov-2010}. Our results are thus in line with the idea that the 
concentration of these membrane nematogens are a major determinant for the amount of ER in sheets or tubules\cite{Kozlov-2010}.

\section{Conclusion}\label{sec:conclusion}

We have described  the membrane  curvature modifying properties of  anisotropic protein inclusions, like the BAR proteins, in terms of an in-plane nematic field.  We have shown that  the  flexibility of the membrane can   promote aggregation and lateral domain formation of these membrane nematogens, even in the absence of self interactions.  These domains can facilitate  shape changes of the membrane.  The equilibrium shapes obtained are strikingly similar to that seen in experiments involving curvature modifying proteins. Prominent structures seen are tubes and discs and coexistence of them. Depending  on the preferred   curvature of the nematogens, a {\it  protein lattice} with helical nematic orientation around the tube is seen.  The properties of this liquid crystalline structure was further  analyzed from a continuum version of the  model and the dependence of the tube radius and   the orientation of the nematic  with respect to the tube axis  was calculated. We also estimate  the thermally induced fluctuation in these quantities and show that they are comparable to what is seen in experiments. In addition we calculate  the persistence  length of the nematogen induced tubes  and show that it is  in the range of experimentally obtained  values. 
 This analysis provides the necessary basis to obtain estimates of model parameters from experiments on coated membrane tubes. At present the available experimental data are very limited. 

 The present modeling provides additional support to the growing notion of the importance of local curvature 
modulating proteins in membrane shape generation in biological cells. Compared to previous modeling of the role of 
membrane proteins inducing directional membrane curvature we have taken into account that membranes are not fully 
decorated, the in-plane interactions between nematogens and the arbitrary membrane shapes with spherical topology.   
The current work focusses only on the membrane interacting part of the protein. Electric charges in BAR-protein  are mostly localized to its membrane facing domain which in turn interact with anionic lipids and enables them to bind strongly to the membrane.  One natural extension of this model is include the  electrostatic interactions through cytosol between proteins moieties protruding out of the membrane.  \\

We emphasize that the main  aim of this work is to show that anisotropic curvature  induced by the inclusions can lead to  aggregation and interesting shape changes.   This is in contrast to the prevailing assumption that explicit protein interactions are essential for aggregation and formation of protein lattices.  Though a quantitative comparison between the predictions of this model and experiments is  not so easy, the model does demonstrate the possibility of generating many biologically relevant shapes of the vesicle by membrane mediated interactions alone.
\section{Acknowledgements} 
The computational work is carried out at HPC facility at IIT-Madras.

\end{document}